\documentclass[useAMS,usenatbib]{mn2e}
\usepackage{graphicx}

\title[X-ray bursts from IGR J17498--2921]{Thermonuclear X-ray bursts from the 401 Hertz accreting pulsar IGR J17498--2921: indication of burning in confined regions}
\author[Chakraborty and Bhattacharyya]{ Manoneeta Chakraborty$^{1}$\thanks{E-mail: manoneeta@tifr.res.in} and Sudip Bhattacharyya$^{1}$\thanks{E-mail:
sudip@tifr.res.in}\\
$^{1}$Department of Astronomy and Astrophysics, Tata Institute
of Fundamental Research, Mumbai 400005, India}

\begin{document}

\date{
}

\maketitle

\label{firstpage}
\begin{abstract}
We use the 2011 {\it Rossi X-ray Timing Explorer} ({\it RXTE}) proportional counter array (PCA)
data of the 401 Hz accreting pulsar and burster IGR J17498--2921 to perform timing
analysis and time-resolved spectroscopy of 12 thermonuclear X-ray bursts.
We confirm previously reported burst oscillations from this source with a
much higher significance (8.8$\sigma$). We notice that the bursts can be divided
into three groups: big photospheric radius expansion (PRE) bursts are about
ten times more luminous than medium bursts, while
the latter are about ten times more luminous than small bursts. 
The PCA field-of-view of these observations contains several known bursters, and
hence some of the observed bursts might not be from IGR J17498--2921.
The oscillations during big bursts at the known pulsar frequency 
show that these bursts were definitely from IGR J17498--2921.
We find that at least several of the other bursts were also likely originated
from IGR J17498--2921.
Spectral analysis reveals that the luminosity differences among various bursts are
primarily due to differences in normalizations, and not temperatures, even
when we consider the effects of colour factor. This shows
burning on a fraction of the stellar surface for those small and medium bursts,
which originated from IGR J17498--2921.
The low values of the upper limits of burst oscillation amplitude for these bursts
suggest a small angle between the spin axis and the magnetic axis.
We find indications of the PRE nature of a medium burst, which likely originated from
IGR J17498--2921. If true,
then, to the best of our knowledge, this is the first time that two PRE bursts
with a peak count rate ratio of as high as $\approx 12$ have been detected from
the same source.
\end{abstract}

\section{Introduction}\label{Introduction}

Periodic intensity pulsations at the neutron star spin frequency, thermonuclear X-ray
bursts, and burst oscillations (brightness variations close
to the spin frequency during the bursts) are observed from some neutron star
low-mass X-ray binaries (LMXBs; \citet{Lambetal2009, StrohmayerBildsten2006}).
Modelling of these features can be useful to measure the neutron star parameters,
and to probe the strong gravity regime (e.g., \citet{Bhattacharyya2010,Psaltis2008}).
However, in order to use these features as tools, one needs to understand them
sufficiently well. Although, their basic properties are well understood, 
there are several outstanding questions. For example, why some of the neutron star 
LMXBs are accreting pulsars (showing periodic pulsations), while others
are not (e.g., \citet{Ozel2009,Lambetal2009}), what creates burning region
asymmetry during decays of some thermonuclear bursts and gives rise to burst
decay oscillations (\citet{Bhattacharyya2010} and references therein), what 
causes the plausible confinement of the burning regions as indicated from
timing analysis (e.g., \citet{Wattsetal2008}), etc. 
Accreting pulsars showing burst oscillations are the ideal sources to 
address these questions. In this paper, we report the results of detailed
timing and time-resolved spectral analyses of the thermonuclear bursts from the
{\it Rossi X-ray Timing Explorer} ({\it RXTE}) proportional
counter array (PCA) field-of-view (FoV) of the
recently discovered accreting pulsar IGR J17498--2921 with neutron star spin
frequency 400.99 Hz \citep{Papittoetal2011}. \citet{Linaresetal2011} reported
burst oscillations from this source. We confirm this feature
with a much higher significance. We also show spectral indication of thermonuclear 
bursts from confined regions on the neutron star surface.

\section{Data Analysis and Results}\label{DataAnalysisandResults}

We analyze all the {\it RXTE} PCA data 
(32 obsIds between August 13 and September 22; 146.496 ks exposure) 
of the 2011 outburst of the accreting 401 Hz pulsar IGR J17498--2921.
Twelve thermonuclear bursts are detected in the entire data
(see \S~\ref{Discussion} for discussions on thermonuclear origin, and
on the sources which could give rise to these bursts).
We carry out timing analysis and time-resolved spectroscopy, mostly using Good-Xenon
data files ($\sim$0.95 $\mu$s time resolution), in order to study the nature
of these bursts. 

\begin{figure*}
\flushleft
\begin{minipage}{0.44\linewidth}
\hspace*{-0.6cm}
\includegraphics*[width=8.5cm]{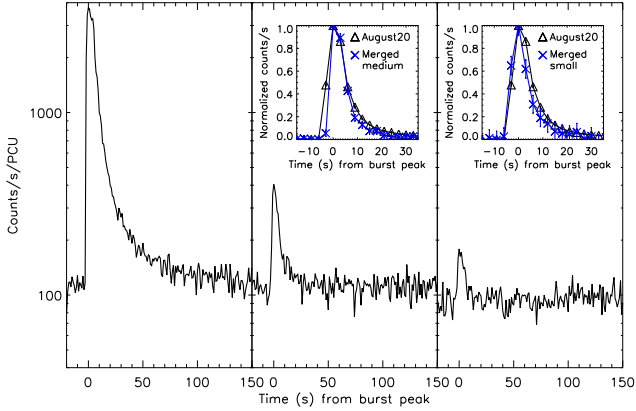}
\caption{Example of each of three types of bursts, big (left panel; August 20, 2011), 
medium (middle panel; August 19, 2011) and small (right panel; August 29, 2011), 
from the IGR J17498--2921 observations. Each {\it RXTE} PCA 
light curve has 1 s time binning. Note that the peak count rate
of the big burst is about an order of magnitude larger than that of the medium burst, and
the latter one is a few times larger than the peak count rate of the small burst. 
In the insets, normalized profile of the August 20 
big burst is compared with that of three combined
medium bursts, and that of seven combined small bursts, after aligning the peaks. These 
insets (with 3 s time bins) show, despite a large change of peak count rate
from one burst to another, the shape/duration of the
big, medium and small bursts are similar to each other (see \S~\ref{DataAnalysisandResults}).
\label{lc}}
\end{minipage}
\end{figure*}

The properties of all the bursts are given in Table~\ref{Properties}.
The bursts can be divided into three groups, based on the 
pre-burst level subtracted peak count rates 
($I_{\rm peak}$: counts per second per proportional counter unit (PCU)): 
two big bursts with $I_{\rm peak} \approx 3600-3700$, three medium bursts with 
$I_{\rm peak} \approx 240-300$, and seven small bursts with
$I_{\rm peak} \approx 45-110$ (see Fig.~\ref{lc}).
However, this figure also shows that the shape and duration of bursts
do not clearly change across the groups.

\begin{figure*}
\flushleft
\begin{minipage}{0.44\linewidth}
\hspace*{-0.5cm}
\includegraphics*[width=8.5cm]{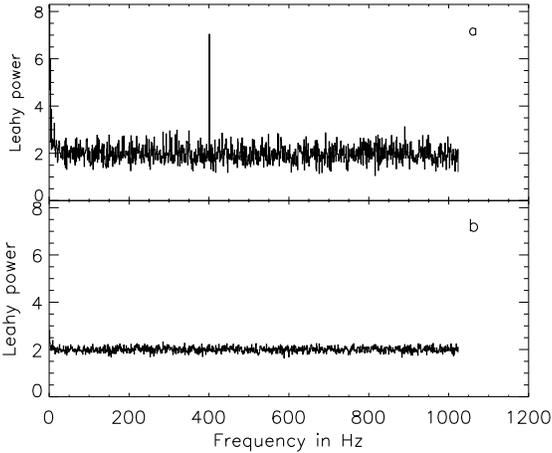}
\caption{Leahy normalized power spectra from the {\it RXTE } PCA data of IGR J17498--2921.
Panel {\it a}: power spectra from 33 time segments (1 s each) during the 
big PRE burst (August 20, 2011) are averaged without any further frequency
rebinning. The strong peak at 401 Hz shows significant burst oscillations.
Panel {\it b}: same as panel {\it a}, but the power spectra from all small and medium bursts 
are merged together. The power spectrum, which is an average of 331 power spectra,
does not show burst oscillations (see \S~\ref{DataAnalysisandResults}).
\label{powerspec}}
\end{minipage}
\end{figure*}

\begin{figure*}
\flushleft
\begin{minipage}{0.44\linewidth}
\hspace*{-0.5cm}
\includegraphics*[width=8.5cm]{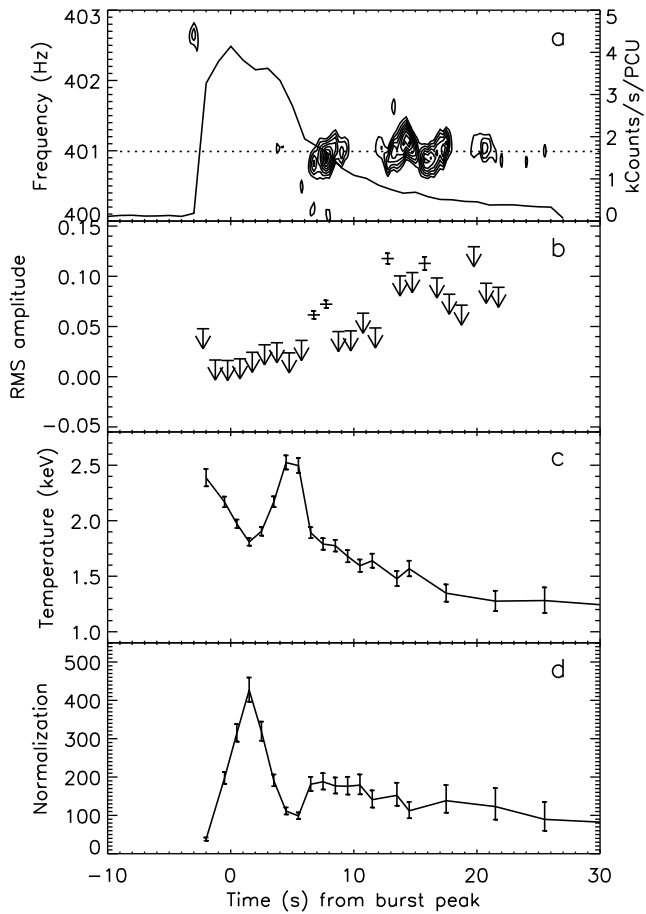}
\caption{Properties of the big burst (August 20, 2011) with oscillations from
IGR J17498--2921.
Panel {\it a}: {\it RXTE} PCA light curve (solid curve) and dynamic power spectrum (contours).
The latter shows 30\% to 99\% contours of the maximum power 37.8, using 2 s overlapping time 
bins with 0.25 s shift between two adjacent bins. These contours suggest the 
presence of an $\approx 401$ Hz signal with no significant 
frequency evolution. The dotted horizontal line shows the pulsar frequency 
(see \S~\ref{DataAnalysisandResults}).
Panel {\it b}: fractional rms amplitudes in 1 s bins during the burst. 1$\sigma$
error bars are given for 3$\sigma$ significant four amplitudes (6.24$\pm$0.39\%, 
7.14$\pm$0.38\%, 11.53$\pm$0.53\% and 11.23$\pm$0.66\%), and 3$\sigma$ upper limits
are given for the rest (see \S~\ref{DataAnalysisandResults}).
Panel {\it c}: the evolution of the best-fit (in $3-15$ keV) blackbody temperature (with 90\%
errors) of the burst (see \S~\ref{DataAnalysisandResults}). The rise of the temperature
is not seen because of the somewhat large time bin.
Panel {\it d}: the evolution of the best-fit (in $3-15$ keV) blackbody normalization (with 90\%
errors) of the burst. This normalization is proportional to the burning area, and is defined as 
$R_{km}^2/D_{10}^2$, where $R_{km}$ is the neutron star radius in km when the 
entire surface emits, and $D_{10}$ is the 
distance to the source in the unit of 10 kpc. The specific correlation between the 
temperature and the normalization shows that this is a PRE burst 
(see \S~\ref{DataAnalysisandResults}).
\label{bigburst}}
\end{minipage}
\end{figure*}

Now we search for oscillations from all the bursts in the entire PCA energy range
(using all active PCUs). We start with the
August 16 big burst, for which burst oscillations were reported (see \S~\ref{Introduction}).
The entire burst (above 5\% of the peak count rate) is divided into 33 ($= M$) segments
of 1 s each. The Leahy normalized power spectra (\citet{Leahyetal1983, vanderKlis1989}) 
from all of them are averaged to obtain a power spectrum of 1 Hz resolution; and a
range of $\pm 3$ Hz from the known pulsar frequency (\S~\ref{Introduction}) is searched
for a candidate peak. This is because the burst oscillation frequency do not
shift from the neutron star spin frequency by more than 3 Hz (e.g., 
\citet{StrohmayerBildsten2006}). We find a candidate peak of $\approx 4.17$ Leahy power at
401 Hz. The probability of obtaining a power this high in a single trial from the
expected $\chi^2$ noise distribution ($2M = 66$ degrees of freedom; \citet{vanderKlis1989})
is $\approx 5.88\times10^{-7}$. Considering a number of trials of 72 ($ = 6\times12$;
six 1 Hz frequency bins are searched for each of 12 bursts), the significance of detection 
is $\approx 4.1\sigma$ (estimated rms amplitude $\approx 4.6\pm0.2$\%).
This suggests that a further and stronger detection would be required for confirmation
of burst oscillations from IGR J17498--2921. Therefore, we perform a similar timing
analysis for the August 20 big burst for $M = 33$. A candidate peak ($\approx 7.04$ Leahy power)
appearing at 401 Hz has the single trial significance of $1-2.24\times10^{-20}$.
Considering a number of trials of 72 (as before), the significance of detection is 
$\approx 8.8\sigma$, which confirms burst oscillations from IGR J17498--2921 (see
Fig.~\ref{powerspec}; note that this burst originated from IGR J17498--2921
(\S~\ref{Discussion})). If some of the 12 bursts were not originated from
IGR J17498--2921 (\S~\ref{Discussion}), then the significance of oscillations from both
the big bursts would be higher.
The dynamic $Z^2$ power spectrum \citep{StrohmayerMarkwardt1999} of the August 20 big burst
shows that the oscillations appear intermittently during burst decay, and there
is no significant frequency evolution (panel {\it a} of Fig.~\ref{bigburst}).
Panel {\it b} of Fig.~\ref{bigburst} shows the rms amplitude evolution 
during the August 20 big burst.
However, burst oscillations are not detected from any medium or small burst. The 3$\sigma$
upper limits of rms amplitude are $4.4$\%, $5.0$\% 
and, $5.1$\%, for all medium and small bursts
combined (Fig.~\ref{powerspec}), for all medium bursts combined, and for all small
bursts combined, respectively.

\begin{figure*}
\flushleft
\begin{minipage}{0.44\linewidth}
\hspace*{-0.5cm}
\includegraphics*[width=8.5cm]{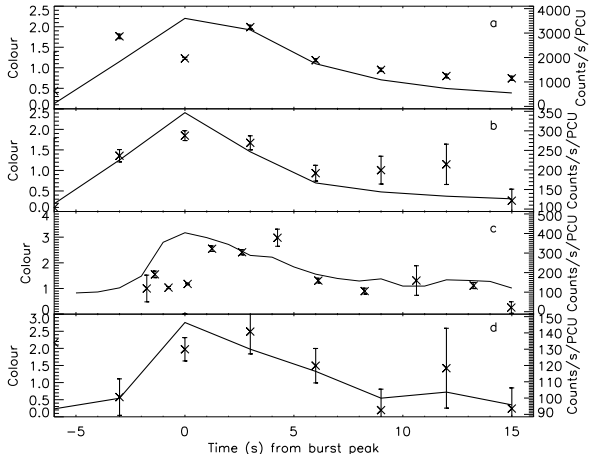}
\caption{The colour (cross sign with 1$\sigma$ error;
ratio of the persistent-subtracted count rate above 6.14 keV to that
below 6.14 kev) and count rate (solid curve)
evolution during the bursts from the {\it RXTE} PCA observations of IGR J17498--2921.
Panel {\it a}: big burst (August 20, 2011).
Panel {\it b}: three medium bursts averaged.
Panel {\it c}: the brightest medium burst (August 19).
Panel {\it d}: seven small bursts averaged.
Burst peaks are made aligned for averaging.
This figure shows that (1) the maximum values of colour for big, medium
and small bursts are similar to each other; (2) there is
a decreasing trend of colour during the decays of
big, medium and small bursts; and
(3) a dip in colour near the peak count rate of the brightest medium burst (August 19)
is consistent with a plausible PRE nature (see \S~\ref{DataAnalysisandResults}).
However, we note that such a dip is not seen in the average profiles (panels
{\it b} and {\it d}).
\label{burstcolor}}
\end{minipage}
\end{figure*}

Next we perform time-resolved spectroscopy of each burst after dividing them into
time segments with sufficient counts. Note that each small burst has just
one segment to maintain enough statistics. From each segment, we create an energy spectrum
with dead time correction \citep{vanderKlis1989}, and a background spectrum from the pre-burst
emission \citep{BhattacharyyaStrohmayer2006,Gallowayetal2008}, considering only the top layers 
of all active PCUs. We fit each energy spectrum in $3-15$ keV with a standard
absorbed blackbody model ({\tt phabs*bbodyrad} in XSPEC; \citet{StrohmayerBildsten2006}) for a
fixed neutral hydrogen column density $N_{\rm H} = 2.87\times10^{22}$ cm$^{-2}$ 
\citep{Torresetal2011}, considering a systematic error of 1\%. The model fits the spectra
well, with reduced $\chi^2 < 1.0$ (degrees of freedom = 28) for $\approx 60$\% spectra, and
between 1.0 and 1.5 for almost all other spectra. The two big bursts show significant cooling
in the decay portions, and the best-fit blackbody temperature 
and normalization (defined in the caption of Fig.~\ref{bigburst}) evolve in
a correlated way. The temperature profile of each big burst shows two peaks, while
a normalization peak coincides with the temperature minimum between two peaks
(see Fig.~\ref{bigburst} for the August 20 burst). This is a clear signature of
a photospheric radius expansion (PRE) burst 
(e.g., \citet{Gallowayetal2008}; see also \citet{Linaresetal2011,
ChakrabortyBhattacharyya2011b}). The medium bursts also show a cooling trend during decay
(e.g., Fig.~\ref{mediumburst}). This figure also shows a somewhat correlation between
the best-fit blackbody temperature and normalization, which is indicative of the PRE nature
of the August 19 medium burst.
(see \S~\ref{Discussion} for a discussion). We find that the best-fit blackbody temperatures of
all the bursts are consistent with each other (see Fig.~\ref{burstspecpar} and \S~\ref{Discussion}).
However, the corresponding best-fit normalizations are correlated with the burst fluence
(integrated energy) values, and both parameters increase roughly by two orders of magnitude
from small bursts to big bursts (see Fig.~\ref{burstspecpar}).
Finally, in order to track the temperature evolution in a model independent way,
we plot the burst colour (ratio of pre-burst level subtracted count rate above 6.14 keV to
that below 6.14 keV) with time (Fig.~\ref{burstcolor}).
This figure supports the finding that the temperatures of all the bursts are consistent 
with each other, and also shows a cooling trend during the decay of small bursts.
A dip in colour (Fig.~\ref{burstcolor}) near the peak count rate supports a plausible PRE nature
of the August 19 medium burst (see above).

\section{Discussion and Conclusions}\label{Discussion}

We discuss the implications of our results in this section.
Let us first show that all the 12 bursts are thermonuclear bursts.
The sharp rise and slow decay of intensity, acceptable fitting of burst
spectra with a blackbody model, cooling during burst decay and detection
of burst oscillations leave no doubt that the big bursts are of thermonuclear
origin. The first three properties are also true for medium and small bursts.
The very similar shape and duration of all the burst profiles argues that, if
big bursts are thermonuclear, then others from the same source are thermonuclear bursts too.
Besides these are not repetitive bursts, and hence are not accretion-powered 
type-II bursts (e.g., \citet{ChakrabortyBhattacharyya2011a}). These establish
that all the 12 bursts are thermonuclear.

The PCA FoV of IGR J17498--2921 contains five additional known thermonuclear
X-ray bursters: XTE J1747--274 (or, IGR J17473--2721), SAX J1750.8--2900,
SAX J1747.0--2853, 1A 1742--289 (or, AX J1745.6--2901) and SLX 1744--299/300.
This brings the question whether some or all of the 12 detected bursts 
(\S~\ref{DataAnalysisandResults}) were not from IGR J17498--2921. The two big
bursts showed oscillations at the known pulsar frequency of 401 Hz, and hence
they were definitely from IGR J17498--2921. But which sources gave 
rise to the medium and small bursts? The first four of the five above-mentioned 
additional bursters are transients, and none of them were in outburst during
the outburst of IGR J17498--2921 (otherwise, the scanning programs of satellites,
such as {\it INTEGRAL} (e.g., \citet{Gibaudetal2011}) 
would detect such an outburst). Since a thermonuclear 
burst from the quiescent phase of a transient is very rare (e.g., \citet{Kuulkersetal2009}),
it is unlikely that the medium and small bursts originated from one of the
four known transient bursters, or any other heretofore unknown transient burster in quiescence in the FoV.
Here we consider that the medium and small bursts have originated from IGR J17498--2921 and/or from the persistent 
burster SLX 1744--299/300 though there still remains a small chance that a low-intensity outburst (missed by X-ray satellites) of a 
transient at the edge of the FOV could give rise one or more of these bursts. Three bursts were previously observed
with PCA from SLX 1744--299/300 \citep{Gallowayetal2008}. We find that the shape and duration 
of these bursts were similar to those of our 12 bursts. The peak count rates (when
corrected for the off-axis position of the source) of all these SLX 1744--299/300
bursts are less than those of our medium bursts, but greater than those of our small bursts. Therefore,
the shape, duration and peak count rate do not reveal whether our small bursts
and medium bursts were from SLX 1744--299/300 or from IGR J17498--2921.
The Table 3 of \citet{Gallowayetal2008} shows that the occurrence of SLX 1744--299/300
bursts per hour was 0.037 for 290 ks of PCA observation. If all our medium and small
bursts had originated from SLX 1744--299/300, then the burst rate of this source would be 
0.246 hour$^{-1}$, which is about seven times larger than the known burst rate 
(0.037 hour$^{-1}$).
Even if only the small bursts had come from SLX 1744--299/300, its burst rate would be
about five times larger than the known rate. These suggest that not all the
medium and small bursts may have originated from SLX 1744--299/300, which implies 
that at least some of
these bursts were from IGR J17498--2921. In order to know whether this is indeed the case,
we attempt to find out if the angular location of a burst is consistent with that 
of IGR J17498--2921. For this, we exploit the fact that the 5 PCUs are not perfectly
aligned \citep{Jahodaetal2006}, and hence the ratio of observed count rates in a pair of 
PCUs depends upon the position of the source within the FoV \citep{Gallowayetal2008}.
Two PCUs were on for each of the bursts, except for the August 16 big burst (1 PCU).
Therefore, for each of 11 bursts, we compute the ratio 
($R_1$, with an error) of observed total counts
(preburst level subtracted and deadtime corrected) in the pair of PCUs.
For the PCA observations of IGR J17498--2921, this source was almost (within a 
few arcsec) at the centre of the FoV. Therefore, if an $R_1$ value is consistent with 
the expected value for the centre of PCA FoV, then the corresponding burst likely
originated from IGR J17498--2921. In order to find out these expected values, we 
consider several bursts in PCA observations 
from two other sources (4U 1636--536 and 4U 1608--52), 
which were at the centre of PCA FoV, and for each of which there is no known burster in the
PCA FoV. For these sources and for each pair of PCUs, we compute a mean burst count ratio
($R_2$, with an error), in the same way we compute $R_1$. These $R_2$'s are the
expected ratio values for the centre of PCA FoV. We compare an $R_1$ value 
with the $R_2$ value for same PCU pair. For each of 11 bursts, 
Table~\ref{Properties} gives the difference between $R_1$ and $R_2$ in the unit 
of the estimated error in this difference. A smaller value of this difference
implies a higher possibility for a burst to be originated from IGR J17498--2921.
For example, this difference is less than one for 
the brightest medium burst (August 19; for which a plausible PRE
nature is seen (\S~\ref{DataAnalysisandResults})) and two small bursts,
implying that these bursts were likely originated from IGR J17498--2921. 

\begin{figure*}
\flushleft
\begin{minipage}{0.44\linewidth}
\hspace*{-0.5cm}
\includegraphics*[width=8.5cm]{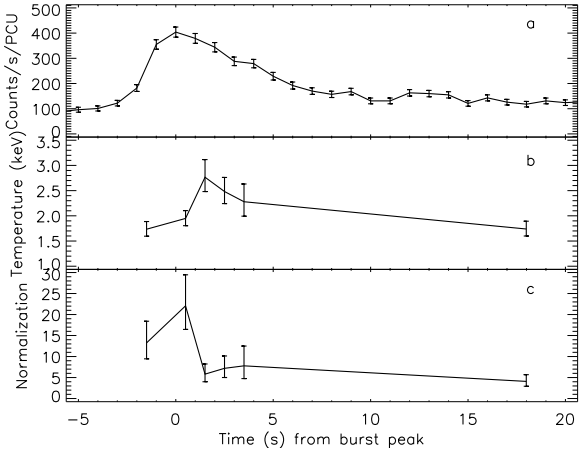}
\caption{Parameter evolution for a medium burst (August 19, 2011) 
from an IGR J17498--2921 observation. 
Panel {\it a}: count rate evolution with 1 s time bins.
Panel {\it b}: the evolution of the best-fit (in $3-15$ keV) blackbody temperature
(with 90\% errors) of the burst.
Panel {\it c}: the evolution of the best-fit (in $3-15$ keV) blackbody normalization
(defined in the caption of Fig.~\ref{bigburst}; with 90\% errors) of the burst.
The specific correlation between the temperature and the normalization
(especially when compared with panels {\it c} and {\it d} of Fig.~\ref{bigburst})
indicates photospheric radius expansion, although the first temperature peak
is not visible (see \S~\ref{DataAnalysisandResults} and \ref{Discussion}).
\label{mediumburst}}
\end{minipage}
\end{figure*}

\begin{figure*}
\flushleft
\begin{minipage}{0.44\linewidth}
\hspace*{-0.5cm}
\includegraphics*[width=8.5cm]{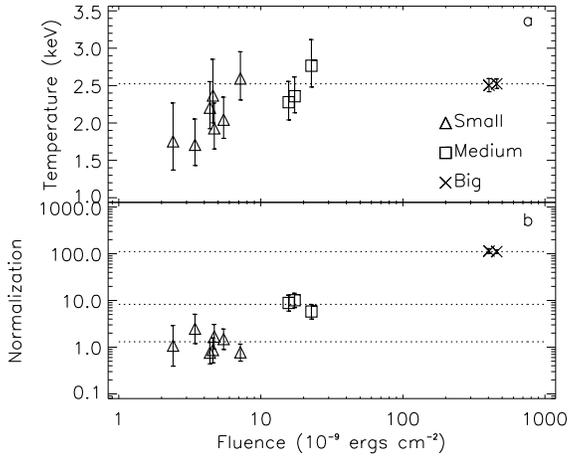}
\caption{Best-fit blackbody temperature (panel {\it a}) 
and normalization (panel {\it b}; 
defined in the caption of Fig.~\ref{bigburst}) in $3-15$ keV as functions of the burst 
fluence for all {\it RXTE} PCA bursts from IGR J17498--2921 observations. 
Note that for each small burst, only one spectrum for the 
entire burst duration is fitted, and the corresponding temperature and
normalization are plotted. For each medium and big burst, the maximum temperature reached during
the burst and its corresponding normalization are plotted.
Cross signs: big bursts; square signs: medium bursts; triangle signs: small bursts.
90\% error bars are given. 
Panel {\it a}: the dotted horizontal line goes through the big burst (August 20, 2011) 
temperature.
Panel {\it b}: the dotted horizontal lines from top to bottom
go through the big burst (August 20, 2011) normalization, mean normalization of medium
bursts, and mean normalization of small bursts respectively.
This figure shows that, while the temperatures of all the bursts are similar to each other,
the normalization (and hence the burning area) is correlated with the burst fluence,
and it increases from small bursts to big bursts by about two orders of magnitude
(see \S~\ref{DataAnalysisandResults}).
\label{burstspecpar}}
\end{minipage}
\end{figure*}

Based on the above discussion and Table~\ref{Properties}, in the
rest of this section we assume that at least one medium burst and at least
one small burst were originated from IGR J17498--2921.
We now ask the question what made some bursts from IGR J17498--2921 
so energetic, while one or more other bursts
were so weak. Was it because of burning of different chemical compositions in
different accretion rate regimes (\citet{StrohmayerBildsten2006} and references therein)?
We find that this is very improbable, because (1) the shape and duration of all the 
burst profiles are similar, indicating the burning of similar compositions; and
(2) the burst fluence is not correlated with the persistent intensity. For example,
the persistent intensities before one of the August 19 small bursts and before the August 29
small burst are respectively slightly higher and slightly lower than the persistent
intensity before the August 20 big burst. Hence we try to find if the increase of
the fluence from small bursts to big bursts for IGR J17498--2921 
is due to the blackbody temperature
increase or due to the burning area ($\propto$ blackbody normalization) increase.
Figs.~\ref{bigburst} ({\it c}), \ref{mediumburst} 
({\it b}), \ref{burstspecpar} ({\it a}) and \ref{burstcolor} show 
that the best-fit temperatures of one burst are consistent with those of others.
In Fig.~\ref{burstspecpar} ({\it a}), the temperatures of small bursts are 
somewhat systematically less than those of medium and big bursts; but this is
because the former temperatures are average values during burst profiles, while
the latter temperatures are maximum values during burst profiles. Note that
such a maximum temperature value for a big burst is for the second temperature peak,
which corresponds to the touch-down point (i.e., settling of the photosphere
on the stellar surface after expansion). Unlike the temperature, the normalization,
which is proportional to the burning area (see caption of Fig.~\ref{bigburst}),
increases significantly with fluence (Fig.~\ref{burstspecpar} ({\it b}); 
\S~\ref{DataAnalysisandResults}). This strongly suggests that the burning area
increases from small bursts to big bursts for IGR J17498--2921, 
while the other parameters, including
the temperature and composition, remain roughly unchanged. If this is true, then each
burst from IGR J17498--2921 
should be similarly strong within its confinement (burning area), and hence,
even the small and medium bursts from this source 
could have local photosphere expansion. However,
for a smaller burst, a signature of such expansion (discussed
in \S~\ref{DataAnalysisandResults}) is washed away due to the large time bins
required to gather sufficient counts for spectral analysis. 
Nevertheless, such a signature appears
for the brightest medium burst (Figs.~\ref{mediumburst} ({\it b}, {\it c}); 
\S~\ref{DataAnalysisandResults}), which likely to be originated from 
IGR J17498--2921 (see Table~\ref{Properties}, burst 3). 
Although, a temperature peak before the 
peak intensity is not there, the following properties indicate
the photosphere expansion: (1) normalization evolution is very similar to that
of the big bursts (compare Fig.~\ref{bigburst} ({\it d}) with Fig.~\ref{mediumburst}
({\it c})); (2) intensity peak corresponds to the normalization peak and a low
temperature; and (3) a temperature peak (corresponding to a low normalization) 
appears after the intensity peak. 
Moreover, an observed dip in colour near the burst intensity peak 
(Fig.~\ref{burstcolor} ({\it c})) 
suggests a similar dip in temperature (undetected plausibly due to large time bin 
as mentioned above), and hence a plausible temperature peak before the peak intensity. However,
a conclusion about the PRE nature of this burst 
has to be made cautiously, because (1) a temperature peak before the 
peak intensity is not significantly detected, and (2) the normalization value following
the maximum is not lower with at least $4\sigma$ significance (as considered
by \citet{Gallowayetal2008}). If this medium burst is really a PRE burst,
then, to the best of our knowledge, this is the first time that two PRE bursts
with a peak count rate ratio of as high as $\approx 12$ have been detected from 
the same source (e.g., \citet{Gallowayetal2006b}).
This makes the standard method of source distance measurement
using PRE bursts (e.g., \citet{Kuulkersetal2003}) somewhat less reliable, 
at least for pulsars. All the bursts from IGR J17498--2921
were likely helium-rich, because of there high temperatures, somewhat short durations
and the PRE nature (for at least the two big bursts; see, for example,
\citet{Gallowayetal2006a, Gallowayetal2008}).

The blackbody normalization values for small and medium bursts 
(Fig.~\ref{burstspecpar} ({\it b})) imply burning areas much smaller than
any realistic neutron star surface area 
(e.g., for a stellar radius of $8-20$ km), implying burning in confined regions. 
Note that these normalization values are estimated assuming a source
distance of 10 kpc (see caption of Fig.~\ref{bigburst}). If the distance is
less (e.g., 7.6 kpc for IGR J17498--2921 
as reported by \citet{Linaresetal2011}), the intrinsic normalization
values will be smaller (implying even smaller burning areas) by a
factor same for all the bursts from a given source. 
A correction due to the surface gravitational
redshift $1+z$ \citep{Sztajnoetal1985} will make the intrinsic normalization values 
further smaller by another factor $(1+z)^2$, which is same for all the bursts
from a given source. 

The absorption and scattering in the neutron star atmosphere makes the observed
temperature higher (relative to the intrinsic temperature) by a factor $f$, 
and the observed normalization lower (relative to the intrinsic normalization) by a
factor $f^4$, where $f$ is the colour factor \citep{Sztajnoetal1985,
Majczynaetal2005,Suleimanovetal2011,Bhattacharyyaetal2010}. Does this mean that
our conclusion about  burning in confined regions for IGR J17498--2921 
is not robust? In order to find out,
let us examine, if the burning areas of the big bursts cover the entire
neutron star surface, then, whether the burning area of a medium
burst from IGR J17498--2921 can also cover the entire surface. 
How could this happen? Suppose, the intrinsic temperature
of the medium burst is smaller than that of big bursts, and $f$ increases with the
decrease of temperature to make the observed temperature of the medium burst and
big bursts similar. Then the ratio of a big burst intrinsic normalization
to the medium burst intrinsic normalization will be lower 
(by a factor, say, $g$) than the ratio of a big burst 
observed normalization to a medium burst observed normalization (the observed 
normalizations are given in Fig.~\ref{burstspecpar} ({\it b})). If $g \sim 10$,
then the burning areas of the medium bursts could be similar to 
those of big bursts (Fig.~\ref{burstspecpar} ({\it b})). 
But, considering the relevant extreme $f$ values 
(1.64, 1.22) from the tables of \citet{Majczynaetal2005}, $g < 3.3 (= [1.64/1.22]^4)$. 
Therefore, even if we consider the maximum possible value (i.e., the
entire surface) for a big burst burning area, a medium burst (from IGR J17498--2921) 
burning area cannot be more than $\sim 35$\% of the stellar surface, even after 
considering the effects of colour factor. With the same arguments, a
small burst (from IGR J17498--2921) burning area will be even smaller. This shows that our 
conclusion about  burning in confined regions is robust.
Finally, we note that the observed burning areas of some of the bursts may be small
due to obscuration (for example, if the observer is close to one spinning pole, and
burning regions are close to the other); but even for this, the burning has to happen
in confined regions. Even if all the medium and small bursts were from
SLX 1744-299/300, the difference in normalization values between these two sets of
bursts would indicate confined burning for small bursts for this source. However,
we note that the confined burning could be more easily explained for
IGR J17498--2921 (than for SLX 1744-299/300), 
which is a known pulsar, and hence likely to have a higher
neutron star magnetic field (see the next paragraph).

Our results support a prediction of \citet{Bildsten1993}, that the weaker bursts are
from small fractions of the neutron stars. The low values of the upper limits of 
burst oscillation fractional rms amplitude 
for small and medium bursts (see \S~\ref{DataAnalysisandResults})
suggest that the burning regions for IGR J17498--2921 
were close to a spinning pole (assuming a hot spot model; \citet{Lambetal2009}).
Hence, a reasonable assumption that the thermonuclear burning happens close to the
magnetic pole(s) (as the accreted matter is channeled to these poles for a pulsar) implies
that the spin axis and the magnetic axis of this pulsar are close to each other.
This supports \citet{Lambetal2009}'s model of accreting millisecond X-ray pulsars.
What gave rise to oscillations during the decays of big bursts? 
If the burning covered the entire neutron star surface, then
the oscillations could be due to a complex asymmetric brightness pattern,
caused by Rossby waves (or r-modes) in the surface layers, shear instabilities, 
etc. (e.g., \citet{LeeStrohmayer2005,Heyl2005,Cumming2005,PiroBildsten2006}). 
Finally, what confined the burning region, and why did the burning area change
from burst to burst? Detailed theoretical and numerical studies are required
to answer these questions. The confined burning could be 
due to the magnetic field near the magnetic pole (but see
\citet{Wattsetal2008}), or, magnetic field locally enhanced by the convection
during thermonuclear flame spreading (e.g., \citet{Spitkovskyetal2002}), or,
magnetic structure (e.g., belt) on the neutron star surfaces 
(e.g., \citet{PayneMelatos2006}),
or, higher local temperature (sufficient for ignition) near the magnetic poles, 
while ignition condition is not reached at rest of the stellar surface 
\citep{Wattsetal2008}, or, something else.
Whatever be the reason, the indication of burning in confined regions reported
in this paper may have significant impact in understanding the accreting pulsars, 
burst oscillations, burst ignition conditions, surface magnetic field structure 
and its interaction with flame spreading, etc. (see also \S~\ref{Introduction}).

\section*{Acknowledgments}

We thank an anonymous referee for very constructive and useful 
comments, which improved the paper.

\begin{table*}
\flushleft
\caption{Properties of bursts from the 2011 {\it RXTE} PCA observations of IGR J17498--2921.}
\begin{tabular}{cccccccc}
\hline
Serial & Observation & Time in&Peak count&$\tau$\footnotemark[4]&Temperature\footnotemark[5]&Normalization\footnotemark[6]&$|R_1-R_2|/\sigma_{(R_1-R_2)}$\footnotemark[7] \\
no. & start time\footnotemark[1] & MJD\footnotemark[2]&rate\footnotemark[3]&  & (keV) & &\\ 
 & & & (count/s) & (s) & & & \\
\hline
 1 & 2011-08-16T15:19:28 & 55789.64 & 3648.72$\pm$61.32 & 6.07 & 2.50$_{-0.09}^{+0.09}$ &  114.35$_{-12.98}^{+14.30}$ &\\
 2 & 2011-08-19T12:11:28 & 55792.53 &  109.04$\pm$14.90 & 4.13 & 2.37$_{-0.37}^{+0.49}$ &    0.87$_{-0.41}^{+0.71}$ & 1.37 \\
 3 & 2011-08-19T13:45:20 & 55792.58 &  291.08$\pm$20.05 & 5.20 & 2.77$_{-0.28}^{+0.35}$ &    5.81$_{-1.81}^{+2.42}$ & 0.65 \\
 4 & 2011-08-19T15:19:28 & 55792.67 &   58.08$\pm$12.96 & 4.17 & 1.71$_{-0.28}^{+0.35}$ &    2.48$_{-1.29}^{+2.58}$ & 1.03 \\
 5 & 2011-08-20T10:08:32 & 55793.43 &   76.69$\pm$13.89 & 4.31 & 1.93$_{-0.28}^{+0.34}$ &    1.69$_{-0.78}^{+1.38}$ & 1.01 \\
 6 & 2011-08-20T10:08:32 & 55793.44 &  242.85$\pm$18.98 & 4.12 & 2.36$_{-0.22}^{+0.26}$ &   10.11$_{-3.07}^{+4.17}$ & 6.89 \\
 7 & 2011-08-20T13:16:32 & 55793.59 & 3634.12$\pm$61.20 & 7.05 & 2.52$_{-0.06}^{+0.06}$ &  111.20$_{-8.82}^{+9.48}$ & 1.72 \\
 8 & 2011-08-21T12:46:24 & 55794.57 &  273.42$\pm$19.47 & 4.17 & 2.28$_{-0.24}^{+0.28}$ &    8.85$_{-2.93}^{+4.16}$ & 1.96 \\
 9 & 2011-08-29T08:47:28 & 55802.39 &   84.51$\pm$13.31 & 3.30 & 2.59$_{-0.29}^{+0.36}$ &    0.78$_{-0.27}^{+0.39}$ & 0.88 \\
10 & 2011-09-03T07:50:24 & 55807.34 &   60.40$\pm$11.96 & 6.20 & 2.04$_{-0.25}^{+0.30}$ &    1.49$_{-0.60}^{+0.95}$ & 1.15 \\
11 & 2011-09-08T05:49:20 & 55812.25 &   45.28$\pm$11.06 & 6.06 & 2.20$_{-0.28}^{+0.35}$ &    0.76$_{-0.32}^{+0.50}$ & 0.26 \\
12 & 2011-09-22T09:25:20 & 55826.41 &   83.99$\pm$11.88 & 3.13 & 1.75$_{-0.38}^{+0.51}$ &    1.08$_{-0.68}^{+1.83}$ & 2.93 \\
\hline
\end{tabular}
\\
$^1$ The start time of the event data file containing the burst.\\
$^2$ The burst peak occurence time in MJD.\\
$^3$ The preburst level subtracted burst peak count rate per PCU (PCU2), 1$\sigma$ error is given.\\
$^4$ The exponential decay time for each burst.\\ 
$^5$ The best-fit blackbody temperature in $3-15$ keV, 90\% error is given (see Fig.~\ref{burstspecpar}).\\
$^6$ The best-fit blackbody normalization in $3-15$ keV, 90\% error is given (see Fig.~\ref{burstspecpar}; defined in the caption of Fig.~\ref{bigburst}).\\
$^7$ $R_1$: ratio of observed total counts of a burst in the pair of PCUs; $R_2$: expected 
value of $R_1$, if the burst were from IGR J17498--2921 (see \S~\ref{Discussion}). 
$\sigma_{(R_1-R_2)}$ is the estimated $1\sigma$ error in $R_1-R_2$. Hence a lower value of
$|R_1-R_2|/\sigma_{(R_1-R_2)}$ implies a higher possibility for a burst to be originated from
IGR J17498--2921. For the burst 1, only one PCU was on.\\
\label{Properties}
\end{table*}

\end{document}